\documentclass{PoS}
\usepackage{slashed,amsmath,amssymb}
\newcommand{\Tr}{{\rm Tr\,\,}}

\title{TEK twisted gradient flow running coupling}

\ShortTitle{TEK twisted gradient flow running coupling}

\author{Margarita Garc\'{\i}a P\'erez\\
        Instituto de F\'{\i}sica Te\'orica UAM/CSIC, Universidad Aut\'onoma de Madrid, 
E-28048-Madrid, Spain\\
        E-mail: \email{margarita.garcia@uam.es}}

\author{Antonio Gonz\'alez-Arroyo\\
Instituto de F\'{\i}sica Te\'orica UAM/CSIC and  Departamento de F\'{\i}sica Te\'orica, C-15,
Universidad Aut\'onoma de Madrid, E-28049-Madrid, Spain \\
        E-mail: \email{antonio.gonzalez-arroyo@uam.es}}

\author{\speaker{Liam Keegan} \\%
PH-TH, CERN, CH-1211 Geneva 23, Switzerland\\
        E-mail: \email{liam.keegan@cern.ch}}

\author{Masanori Okawa\\
Graduate School of Science, Hiroshima University, 
Higashi-Hiroshima, Hiroshima 739-8526, Japan\\
        E-mail: \email{okawa@sci.hiroshima-u.ac.jp}}

\abstract{We measure the running of the twisted gradient flow coupling in the Twisted Eguchi-Kawai (TEK) model, 
the SU($N$) gauge theory on a single site lattice with twisted boundary conditions in the large $N$ limit.}

\FullConference{The 32nd International Symposium on Lattice Field Theory\\
                 23-28 June, 2014\\
                 Columbia University New York, NY}

\begin{document}

\section{Introduction}
The Twisted Eguchi-Kawai (TEK) model~\cite{GonzalezArroyo:1982hz,GonzalezArroyo:2010ss} provides a single site 
formulation of SU($N$) lattice gauge theory, which in the infinite $N$ limit at fixed bare 't Hooft
coupling reproduces the infinite volume theory. Moreover, the corrections to 
the TEK model at finite $N$ take the form  (at least in perturbation
theory) of finite volume corrections for  an effective lattice size of $\sqrt{N}$.
For example, propagators are identical to those of a normal $(\sqrt{N})^4$ 
lattice~\cite{GonzalezArroyo:1982hz,Perez:2013zna}. If this relation
holds in the continuum limit, it would be possible to use the rank of
the gauge group as a size parameter and determine the running of the
coupling with respect to it. Essentially, we  could simply use a
standard finite volume step scaling procedure and apply it to the
single-site model replacing the linear size of the finite volume,
$l=La$, by $\tilde{l}=\sqrt{N}a$. Usually  the renormalization scale is defined
in terms of the linear size of the finite volume, $l=La$, and the
change of scale $l\rightarrow sl$ is accomplished by changing the number of points in the lattice,
$L\rightarrow sL$. For the TEK model, the lattice always consists of a single
site, and we implement the change of scale  $\tilde{l}\rightarrow
s\tilde{l}$ by  scaling the rank of the gauge group, SU($N$) $\rightarrow$ SU($s^2 N$).
To extract the continuum renormalized coupling
constant, one should take the $a\longrightarrow 0$ ($N \longrightarrow
\infty$) limit. Thus,  according to volume independence,  we expect
that the renormalized coupling coincides with that of the ordinary
pure gauge theory  at  $N=\infty$. 

To be specific, we based our method on a finite volume running coupling scheme defined 
using the gradient flow~\cite{Luscher:2010iy} in a four dimensional torus with twisted boundary 
conditions in one plane~\cite{Ramos:2013gda,Ramos:2014kla}.
Here we define an analogous scheme in the TEK single site model, 
which has twisted boundary conditions in all directions, and use it to perform a step scaling study
and determine the running of the coupling in SU($N$) gauge theory in the large $N$ limit.

\section{TEK Volume Independence}
\begin{figure}
 \begin{center}
 \includegraphics[width=5cm,angle=270]{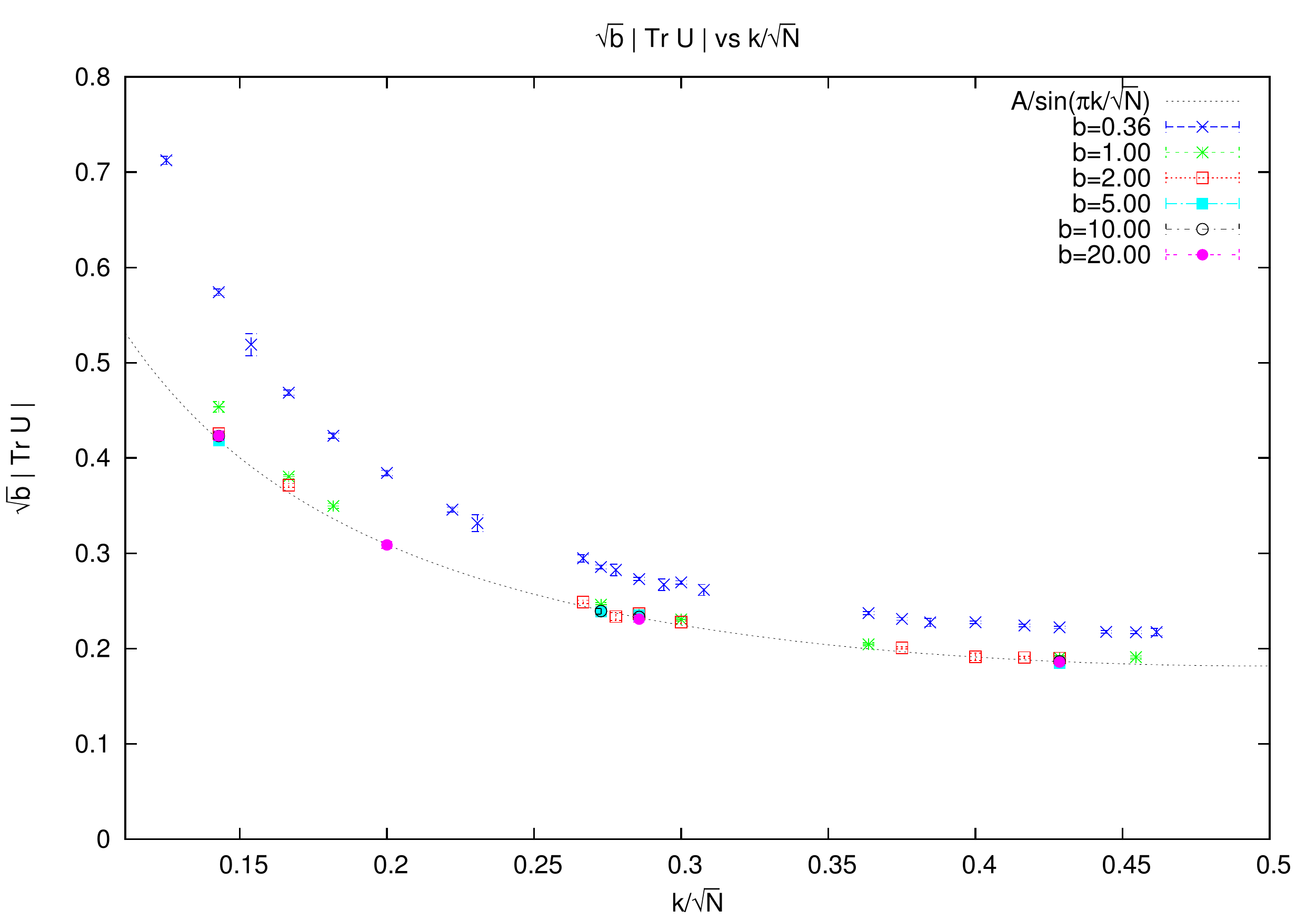} \includegraphics[width=5cm,angle=270]{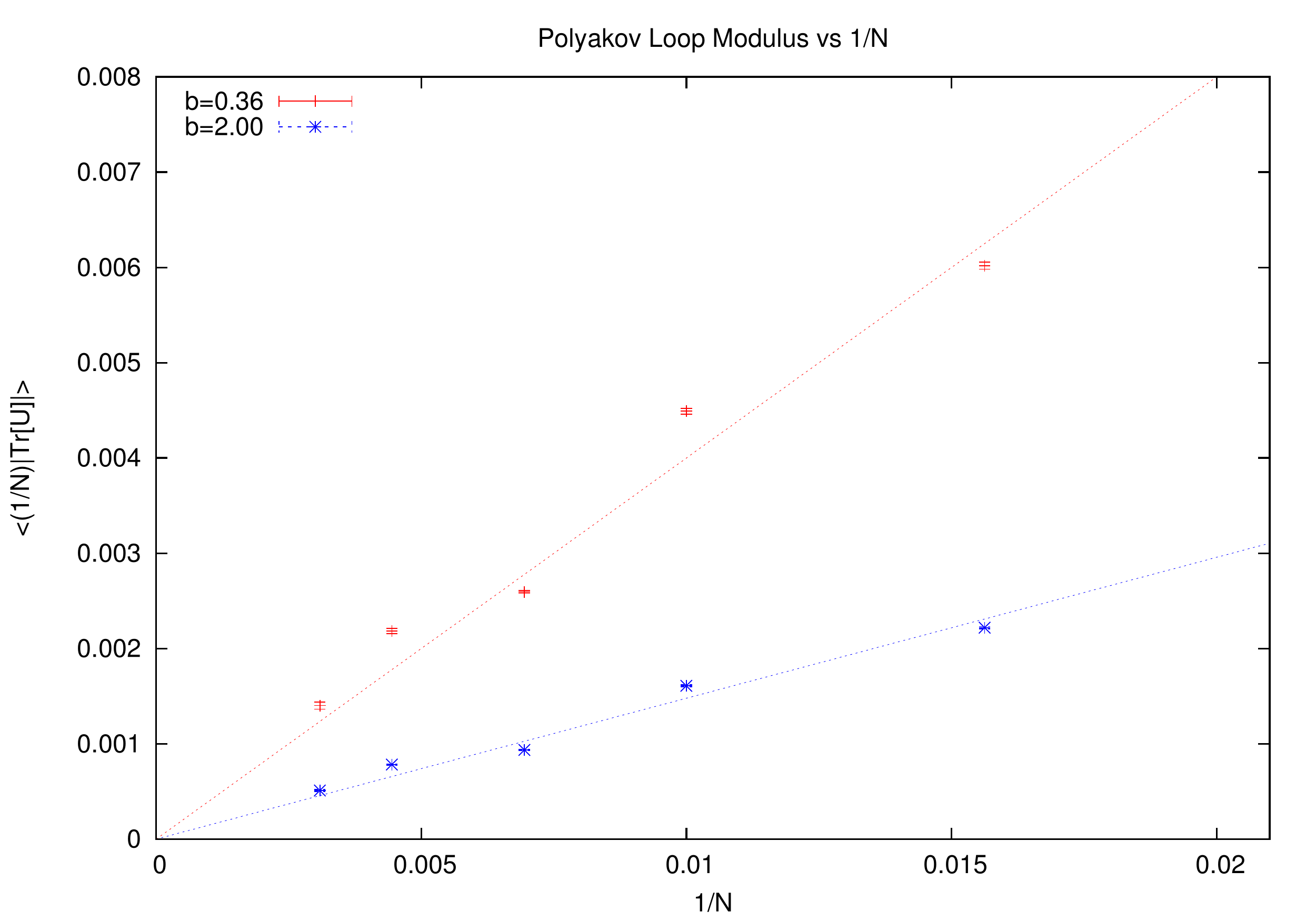}
 \caption{
\label{fig:poly}
Left: $\sqrt{b}\left| \Tr U_{\mu} \right|$ vs $k/\sqrt{N}$ compared to the pertubative expectation, for many values of $b$ and $N$. Right: $\tfrac{1}{N}\left| \Tr U_{\mu} \right|$ vs $1/N$ at $b=0.36$ and $b=2.00$. For all values of the coupling this quantity seems to go to zero in the large $N$ limit. The scatter of the points is caused by $k/\sqrt{N}$ varying somewhat with $N$.}
 \end{center}
\end{figure}

The TEK model consists of 4 SU($N$) matrices $U_{\mu}$, with the action
\begin{equation}
 S = bN \sum_{\mu\nu}\left(N - z_{\mu\nu} \Tr \left[U_{\mu}U_{\nu}U^{\dagger}_{\mu}U^{\dagger}_{\nu}\right] \right), \quad z_{\mu\nu} = z^{*}_{\nu\mu} = e^{2\pi i k/\sqrt{N}}\,\,\mathrm{for}\,\,\mu<\nu,
\end{equation}
where $b$ is the lattice analog of the inverse 't Hooft coupling, $1/(Ng^2)$, and the flux $k$ is 
an integer coprime with $\sqrt{N}$. In order for volume independence to hold in the large $N$ limit, 
center symmetry must not be spontaneously broken, i.e. the trace of all open Wilson loops on the lattice
should go to zero in this limit. This is the case if the flux $k$ is chosen to satisfy $k/\sqrt{N} > 1/9$.
The left--hand plot of Fig.~\ref{fig:poly} shows the quantity $\sqrt{b} \left| \Tr U_{\mu} \right|$ as a 
function of $k/\sqrt{N}$ for many values of $N$ and $b$, along with the perturbative prediction,
$\sqrt{b}\left| \Tr U_{\mu} \right| \propto 1/\sin(\pi k/\sqrt{N})$, which is in good agreement with the
data for $b\gtrsim 2.0$. This quantity divided by $N$ is shown as a function of $1/N$ for the weakest and
strongest values of the coupling used for the step scaling analysis in the right--hand plot of Fig.~\ref{fig:poly}. Since
this goes to zero in the large $N$ limit, reduction should hold.

To all orders in perturbation theory, the TEK model is equivalent to
ordinary  lattice gauge theory, up to corrections that depend on the parameter $\tilde \theta$,
where $\tilde\theta = 2\pi \bar k/\sqrt{N}$, and $\bar k$ is defined as the integer that satisfies
$k \bar k = 1\, (\mathrm{mod}\,\, \sqrt{N})$. To eliminate this effect
one should  scale $k$ and $\sqrt{N}$ so as to keep $\tilde \theta$ constant
in  the large $N$ limit. Strictly speaking this is not possible as
$\bar k$ and $\sqrt{N}$ have to be coprime. This is a source of
systematic error in our results.  However, the variations in $\tilde
\theta$ can be made smaller  for larger values of $N$, since there are more possible choices for $k$. The values of $k$ and $N$ used
in this work are listed in Tab.~\ref{tab:k}. 

\section{Gradient Flow Coupling}
At positive Yang--Mills gradient flow time, the action density of SU($N$) gauge theory is a renormalized 
quantity which, in infinite volume, has  a perturbative expansion~\cite{Luscher:2010iy},
\begin{equation}
\left\langle E(t) \right\rangle=\tfrac{1}{4} \left\langle G^a_{\mu\nu}(t)G^a_{\mu\nu}(t) \right\rangle = \frac{3(N^2-1)}{128 \pi^2 t^2}g^2_{\overline{MS}} + \mathcal{O}(g^4_{\overline{MS}}).
\end{equation}
If we work at finite box size  $l$ and fix the flow time to a constant fraction of this size, $\sqrt{8t}=cl$, we obtain   a quantity that
depends on a single length scale and that to leading order is proportional
to the coupling constant. This allows us to define a 
renormalized coupling as follows~\cite{Ramos:2013gda,Ramos:2014kla}:
\begin{equation}
\label{coupling}
\lambda_{TGF}(l) \equiv \left. \mathcal{N}_T^{-1}(c) \,
\tfrac{1}{N}t^2\langle E \rangle \right|_{t=c^2 l^2/8} =
\lambda_{\overline{MS}} + \mathcal{O}(\lambda^2_{\overline{MS}})
\end{equation}
where $\lambda = Ng^2$  is the `t Hooft coupling. The constant
$\mathcal{N}_T(c)$  is a kinematic factor that ensures that to leading
order $\lambda_{TGF}=\lambda_{\overline{MS}}$. In the previous
definition  the constant $c$ is kept fixed as the scale is changed. 
A change in $c$ can be considered a change of renormalization scheme. 

In this work, we use this idea  to
construct the equivalent renormalized coupling definition for the TEK model 
where $\tilde l=a \sqrt{N}$ replaces $l$. We also need to define the 
observable that  will be used to estimate the action density. The simplest
choice is the plaquette $E_P$:
\begin{equation}
E_P=\sum_{\mu\nu}\left(N-z_{\mu\nu}\Tr\left[U_{\mu}U_{\nu}U_{\mu}^{\dagger}U_{\nu}^{\dagger}\right]\right)
\end{equation}
It is convenient to adjust $\mathcal{N}_T(c)$  to preserve the
equality of the bare and renormalized  couplings at leading order on the lattice.
This gives
\begin{equation}
\mathcal{N}_T^P(c) = \frac{3c^4}{128} \sum'_{n} e^{- c^2 N \sum_{\rho}
\sin^2(\pi n_{\rho}/\sqrt{N})}.
\end{equation}
where $n_{\rho}=0,1,\dots,\sqrt{N}-1$ and the prime in the sum means
that we do not include the term with all $n_{\rho}=0$.

Alternatively,  one can take a different observable $E_S$, which we will call
{\em symmetric},  as follows
\begin{equation}
E_S=-\tfrac{1}{128}\sum_{\mu,\nu}\Tr \left\{\left[
z_{\nu\mu} U_{\nu}U_{\mu}U_{\nu}^{\dagger}U_{\mu}^{\dagger} 
+ z_{\nu\mu} U_{\mu}U_{\nu}^{\dagger}U_{\mu}^{\dagger}U_{\nu} + 
z_{\nu\mu} U_{\nu}^{\dagger}U_{\mu}^{\dagger}U_{\nu}U_{\mu} + 
z_{\nu\mu} U_{\mu}^{\dagger}U_{\nu}U_{\mu}U_{\nu}^{\dagger} -
h.c.\right]^2\right\}.
\end{equation}
We can compute the corresponding factor $\mathcal{N}_T(c)$ which is given by
\begin{equation}
\mathcal{N}_T^S(c) = \tfrac{c^4}{512} \sum_{\mu\neq\nu}\sum_{n} e^{-
c^2 N \sum_{\rho} 
\sin^2(\pi n_{\rho}/\sqrt{N})}\frac{\sin^2(2\pi n_{\mu}/\sqrt{N})
\cos^2(\pi n_{\nu}/\sqrt{N})}{\sum_{\rho} 
\sin^2(\pi n_{\rho}/\sqrt{N})}.
\end{equation}
We emphasize that choosing $\mathcal{N}_T$ for each observable as we
have done considerably reduces lattice artefacts as compared to using 
the continuum factor 
\begin{equation}
\mathcal{N}_T(c) = \tfrac{3c^4}{128} \sum_{n\in Z^4-\{0\}} e^{-\pi^2 c^2 n^2} =
\tfrac{3c^4}{128} \left[(\theta_3(0,i \pi c^2))^4 - 1 \right]
\end{equation}
where the sum is over all non-zero integer 4-vectors, and
$\theta_3(0,z)$ is the Jacobi function.

We define a continuum step scaling function in the usual way:
\begin{equation}
\sigma(u,s) = \left.\lambda_{TGF}(s\tilde{l})\right|_{\lambda_{TGF}(\tilde{l})=u}
\end{equation}
In order to obtain this quantity from the TEK model, one starts with the
lattice equivalent 
\begin{equation}
\Sigma(u,s,\sqrt{N}) = \left.\lambda_{TGF}(s\sqrt{N},b)\right|_{\lambda_{TGF}(\sqrt{N},b)=u}
\end{equation}
and then take the continuum limit ($a \longrightarrow  0
\Leftrightarrow  N\longrightarrow \infty$) keeping
$u$ fixed.

\section{Numerical Determination of $\lambda_{TGF}$}

We choose a step scaling factor $s=3/2$, and simulate a series of pairs of $N$; 
$\sqrt{N}=8,10,12$ and $s\sqrt{N}=12,15,18$. The corresponding run parameters are 
listed in Table~\ref{tab:k}. For each $N$ we simulate a series of bare couplings 
going from weak ($b=2.00$) to strong ($b\simeq0.36$) coupling. Each configuration 
generated in the simulation is separated by a number of sweeps, where each sweep 
consists of one heat--bath and 5 over--relaxation updates. The number of sweeps 
is chosen such that autocorrelations are negligible. This number increases both 
with $N$ and as the coupling is made stronger, and goes up to 1600 sweeps between
each measurement for $N=324$ at $b=0.37$. The Wilson flow is integrated using
the 3rd order Runge--Kutta scheme proposed in Ref.~\cite{Luscher:2010iy}, choosing
the integration stepsize between $0.01$ and $0.03$, and ensuring that the resulting
integration errors are much smaller than the statistical uncertainties.

\begin{table}
\begin{center}
\begin{tabular}{c|c|c|c|c|c}
& $\sqrt{N}=8$ & $\sqrt{N}=10$ & $\sqrt{N}=12$ & $\sqrt{N}=15$ & $\sqrt{N}=18$\\
\hline
$(k, \bar k)$ & (3,3) & (3,3) & (5,5) & (4,4) & (5,7)\\
$\tilde\theta/2\pi = \bar k / \sqrt{N}$ & 0.375 & 0.300 & 0.417 & 0.267 & 0.389 \\
\end{tabular} 
\end{center}
\caption{Run parameters for each $N$, where $k$ is the flux, and $\tilde\theta/2\pi = \bar k / \sqrt{N}$
is the quantity we would ideally keep constant for all $N$.}
\label{tab:k}
\end{table}

The parameter $c$ is in principle arbitrary, and different values correspond to different renormalization schemes. 
In general, a smaller value of $c$ will result in smaller statistical uncertainties, but at the cost of larger 
lattice artefacts, and vice versa. Here we take $c=0.30$ as a good compromise between these two effects. 
The measured couplings using the symmetric definition are listed in Tab.~\ref{tab:E}, and have statistical 
errors $\mathcal{O}(0.3-0.5\%)$.

\begin{table}
\begin{center}
\scriptsize
\begin{tabular}{c|c|c|c|c|c}
$b$ & $\sqrt{N}=8$ & $\sqrt{N}=10$ & $\sqrt{N}=12$ & $\sqrt{N}=15$ & $\sqrt{N}=18$ \\
\hline
0.360 & 16.643(77) & 21.05(10) & 25.60(12) & - & - \\
0.365 & 14.383(61) & 17.492(82) & 20.755(88) & - & - \\
0.370 & 12.979(53) & 15.445(67) & 17.857(81) & 23.52(11) & - \\
0.375 & 11.843(45) & 13.672(58) & 15.698(63) & 19.788(94) & 24.17(11) \\
0.380 & 10.986(40) & 12.469(51) & 14.051(57) & 17.350(81) & 20.496(97) \\
0.390 & 9.624(33) & 10.685(40) & 11.801(44) & 13.882(59) & 15.626(66) \\
0.400 & 8.601(28) & 9.402(33) & 10.246(37) & 11.579(44) & 13.001(54) \\
0.420 & 7.091(22) & 7.652(25) & 8.190(28) & 8.966(29) & 9.727(35) \\
0.450 & 5.718(17) & 6.075(19) & 6.351(19) & 6.796(20) & 7.185(24) \\
0.500 & 4.370(12) & 4.546(13) & 4.726(14) & 4.971(14) & 5.156(15) \\
0.600 & 2.9878(76) & 3.0635(79) & 3.1536(87) & 3.2498(86) & 3.3256(88) \\
0.800 & 1.8556(46) & 1.8815(48) & 1.9041(47) & 1.9419(47) & 1.9720(49) \\
1.000 & 1.3434(33) & 1.3618(32) & 1.3747(33) & 1.3990(35) & 1.4123(36) \\
1.200 & 1.0603(26) & 1.0712(26) & 1.0747(26) & 1.0842(26) & 1.0990(27) \\
1.500 & 0.8030(19) & 0.8101(20) & 0.8127(20) & 0.8227(20) & 0.8255(20) \\
2.000 & 0.5716(13) & 0.5752(13) & 0.5771(14) & 0.5805(13) & 0.5826(14)
\end{tabular}
\end{center}
\caption{Measured coupling $\lambda_{TGF}$ for each $b$ and $N$ (symmetric definition).}
\label{tab:E}
\end{table}

\begin{figure}
 \begin{center}
 \includegraphics[width=7.5cm,angle=0]{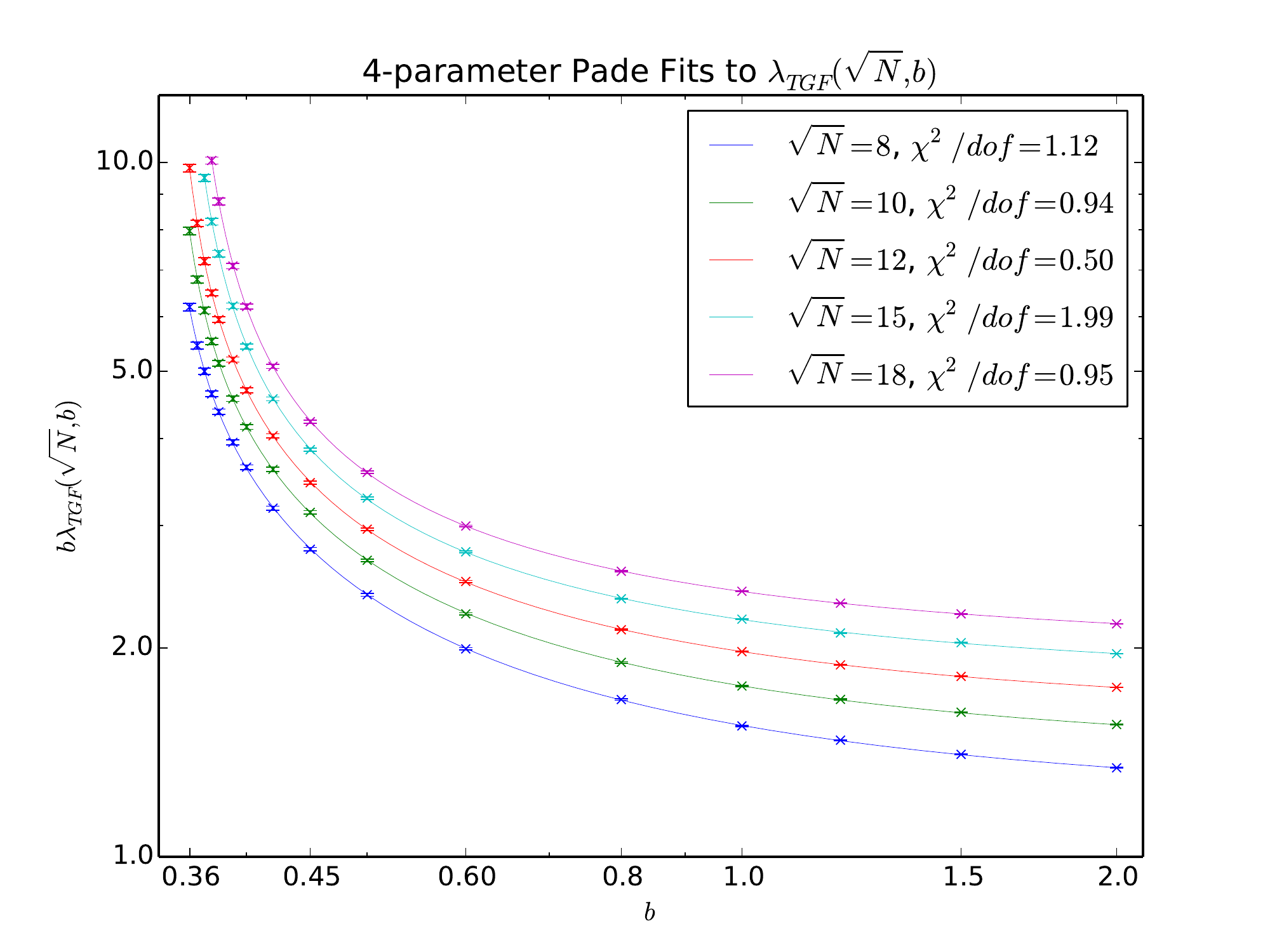}\includegraphics[width=7.5cm,angle=0]{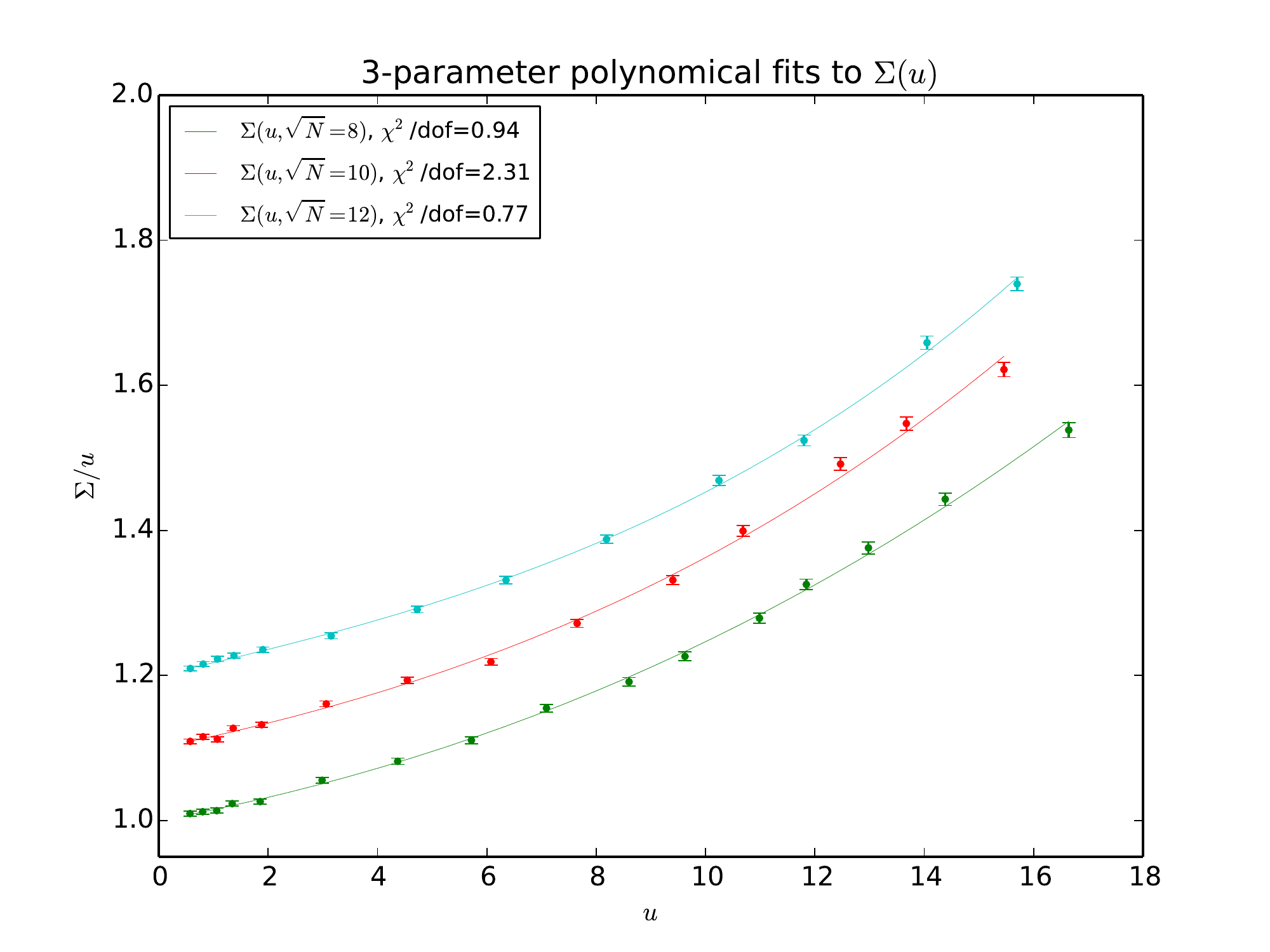}
 \caption{
 \label{fig:interpolate}
 To take the continuum limit at fixed $u$ we need to interpolate the data, which is done in two ways. 
 Left: 4--parameter Pad\'e interpolation of $\lambda_{TGF}(\sqrt{N},b)$ in $b$ (data at different $N$ 
 displaced vertically by 0.2). Right: 3--parameter polynomial interpolation of $\Sigma(u,s,\sqrt{N})/u$ 
 in $u$ (data displaced vertically by 0.1).}
 \end{center}
\end{figure}

\section{Continuum Extrapolation of the Step Scaling Function}
\begin{figure}
 \begin{center}
 \hspace{-0.8cm}\includegraphics[width=8.3cm,angle=0]{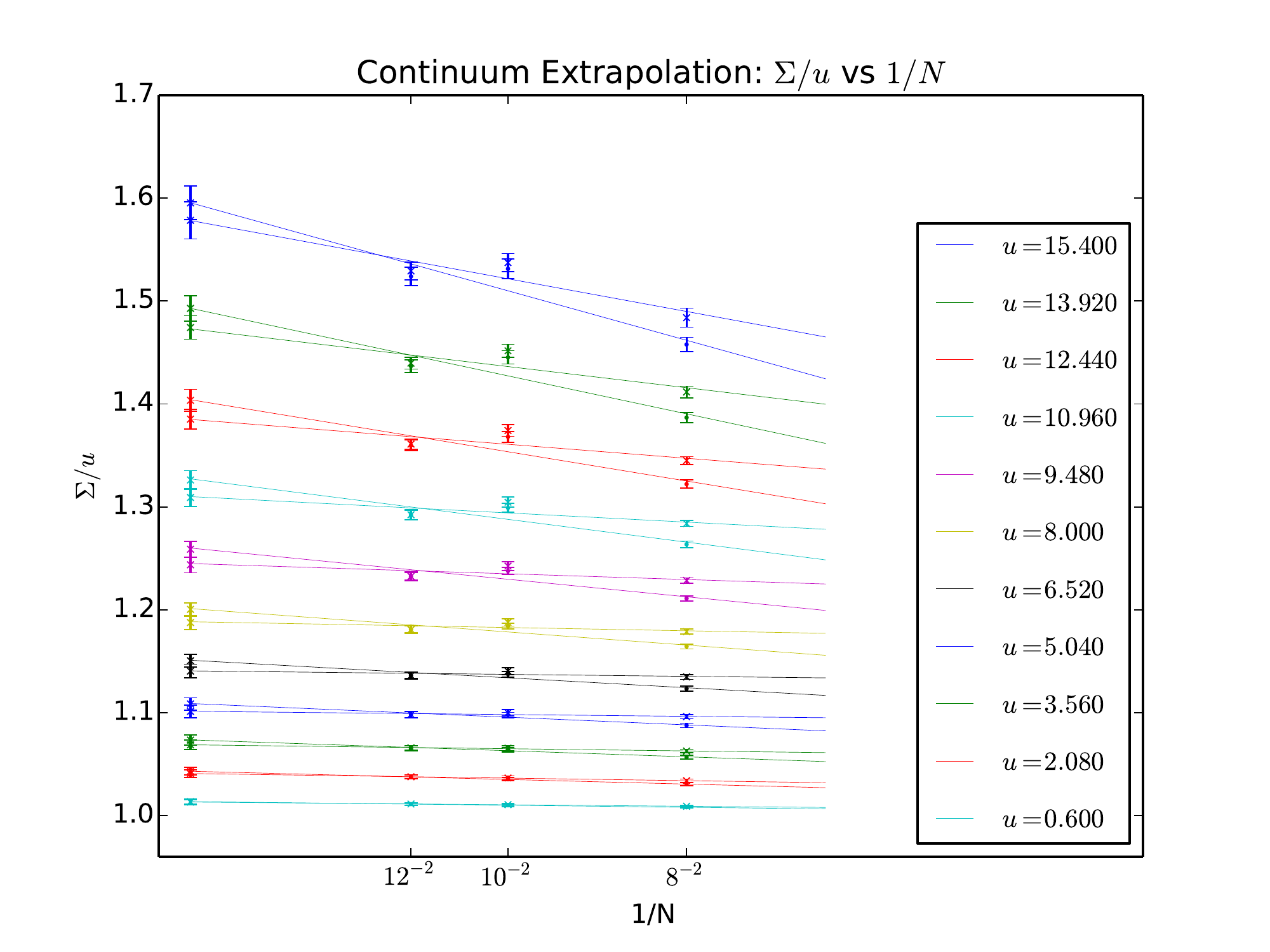} \hspace{-0.9cm} \includegraphics[width=8.3cm,angle=0]{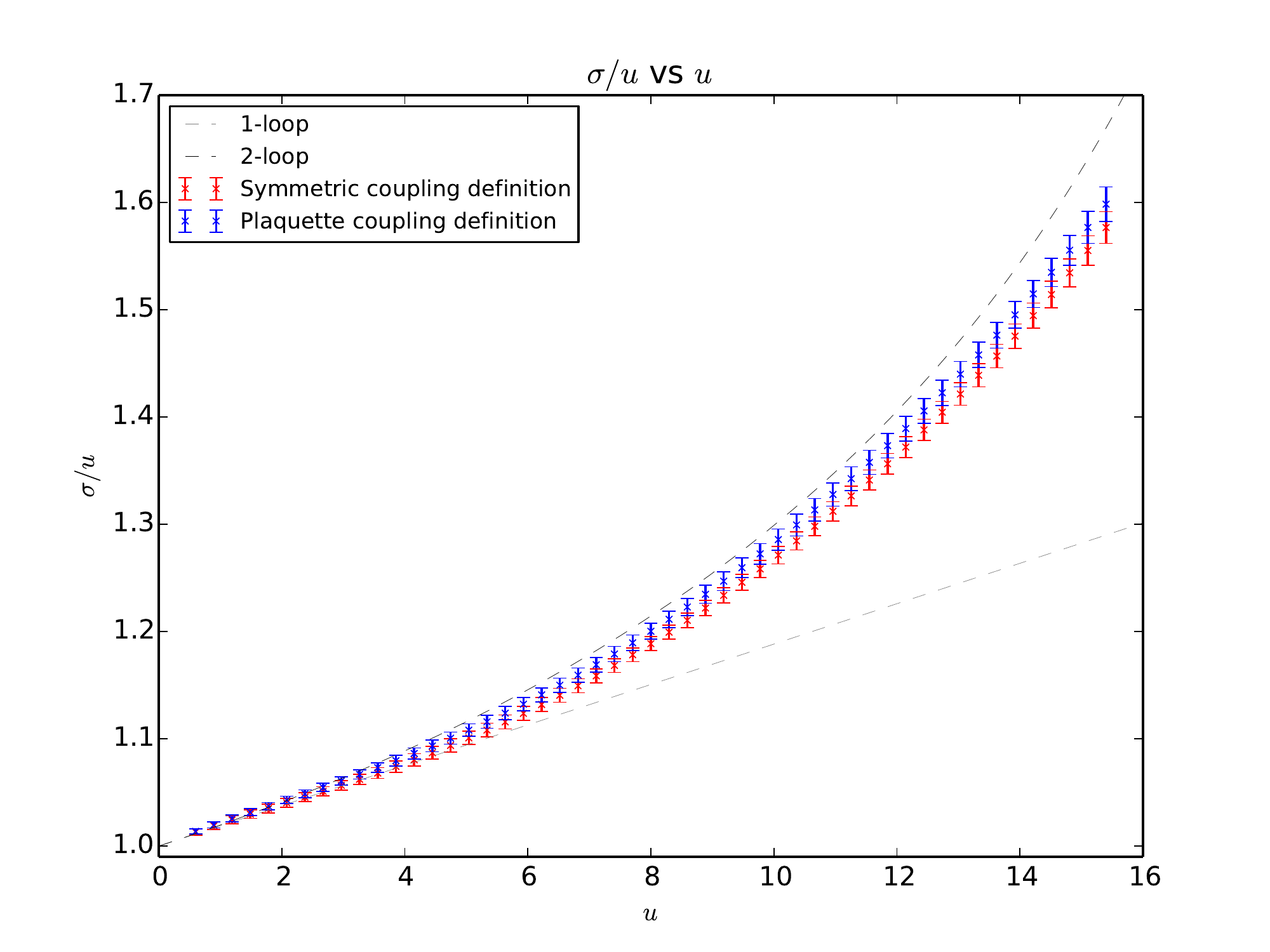} \hspace{-0.9cm}
 \caption{
 \label{fig:extrap}
 Left: Some examples of the continuum extrapolation of $\Sigma(u,\sqrt{N})/u$ in $1/N$, using both the plaquette (points) and symmetric (crosses) definitions, which extrapolate to continuum values that are 
 consistent within errors.
 Right: The final continuum determination of $\sigma(u,s=3/2)/u$ vs $u$, using the plaquette (blue) and symmetric (red) definitions, along with the 1--loop and 2--loop perturbative predictions.
 The agreement between our data at weak coupling and the perturbative prediction is very good.
 }
 \end{center}
\end{figure}

To take the continuum limit of the step scaling function at fixed $u$ we need to interpolate the data. 
In order to check for systematics we do this in two different ways. The first is to interpolate the coupling
as a function of $b$, for each $N$, using a 4--parameter Pad\'e function of the form
\begin{equation}
 \lambda_{TGF}(N,b) = \frac{1}{b}\frac{a_0 + a_1 b + b^2}{a_2 + a_3 b + b^2},
\end{equation}
as proposed in Ref.~\cite{Ramos:2013gda}. This allows us to determine the coupling for each $N$ at any value of $b$, 
and hence the step scaling function at any value of $u$. The second interpolation strategy is to first 
construct the lattice step scaling function directly from the data for the available values of the coupling,
then to interpolate this quantity as a function of $u$ using a 3--parameter polynomial of the form
\begin{equation}
\Sigma(u,\sqrt{N})/u = 1 + a_0 u + a_1 u^2 + a_2 u^3.
\end{equation}
Both these fit functions are constructed to have the correct leading order behaviour in the weak coupling limit, 
i.e. $\lambda_{TGF}(N,b)\rightarrow 1/b$ and $\Sigma(u,\sqrt{N})\rightarrow u$. Examples of both fits are shown
in Fig.~\ref{fig:interpolate}, where the data at different $N$ have been displaced vertically for clarity, and 
have a similar $\chi^2/\mathrm{d.o.f.} \sim 1$. Some examples of the resulting continuum extrapolation in $1/N$
are shown in the left hand plot of Fig.~\ref{fig:extrap}. At each value of $u$ there are two separate continuum extrapolations. 
The data for the symmetric definition of the coupling are shown as crosses, while those using the plaquette 
definition are shown as points. The difference between the two definitions at finite $N$ is a measure of the 
size of lattice artefacts, and the two definitions should extrapolate to consistent values in the continuum limit. 
The errorbars are determined using bootstrap replicas of the data, and using both interpolation strategies, so they
include both the statistical errors and the systematic errors due to the interpolation. They do not however include the
systematic error due to $\tilde \theta$ not being kept exactly constant as we take the continuum limit. 
Indeed the fact that the $\sqrt{N}=10$ points are systematically higher than those at other values of $N$ in the extrapolations
is presumably due to this effect.

The final continuum determination of $\sigma(u)/u$ is shown in the right hand plot of Fig.~\ref{fig:extrap} as a function of $u$,
along with the 1--loop and 2--loop perturbative predictions.

\section{Conclusion}

We define a scale-dependent renormalized coupling constant for the 
SU(N) single-site  TEK model by replacing  the space-time  size parameter of certain
definitions by an effective size determined uniquely by the rank $N$ of the
group.  We use standard methods to determine the running of the
coupling and the  step-scaling function over a wide range of scales. 
The lattice step-scaling is extrapolated to the continuum limit by
taking the $N\longrightarrow \infty$ limit at fixed values of the
coupling. An optimal extrapolation should be done keeping
$\tilde{\theta}=2\pi \bar k/\sqrt{N}$ approximately constant. This is a source of systematic 
errors which does not seem to have a strong impact on the result. 

The resulting extrapolated step-scaling function shows a
similar behaviour to standard $SU(N)$ definitions and matches the
perturbative prediction at weak coupling. This result provides evidence 
that the relation between finite rank and finite volume is preserved
in the continuum limit.

\appendix
\section*{Acknowledgments}
We acknowledge financial support from the MCINN grants FPA2012-31686 and FPA2012-31880, the Comunidad Aut\'onoma 
de Madrid under the program HEPHACOS S2009/ESP-1473, and the Spanish MINECO's ``Centro de Excelencia Severo 
Ochoa'' Programme under grant SEV-2012-0249. M. O. is supported by the Japanese MEXT grant No 26400249. 
Calculations have been done on the Hitachi SR16000 supercomputer both at High Energy Accelerator Research 
Organization(KEK) and YITP in Kyoto University, and the HPC-clusters at IFT.  Work at KEK is supported by the
Large Scale Simulation Program No.13/14-02.

\end{document}